# Effect of the geometry of butt-joint implant-supported restorations on the fatigue life of prosthetic screws

Mikel Armentia, MSc,[a] Mikel Abasolo, PhD,[b] Ibai Coria, PhD,[c] and Nicolas Sainitier, PhD[d]

## ABSTRACT

**Statement of problem.** Dental implant geometry affects the mechanical performance and fatigue behavior of butt-joint implant-supported restorations. However, failure of the implant component has been generally studied by ignoring the prosthetic screw, which is frequently the critical restoration component.

**Purpose.** The purpose of this in vitro study was to evaluate the effect of 3 main implant geometric parameters: the implant body diameter, the platform diameter, and the implant-abutment connection type (external versus internal butt-joint) on the fatigue life of the prosthetic screw. The experimental values were further compared with the theoretical ones obtained by using a previously published methodology.

**Material and methods.** Four different designs of direct-to-implant dental restorations from the manufacturer BTI were tested. Forty-eight fatigue tests were performed in an axial fatigue testing machine according to the International Organization for Standardization (ISO) 14801. Linear regression models, 95% interval confidence bands for the linear regression, and 95% prediction intervals of the fatigue load-life (F-N) results were obtained and compared through an analysis of covariance (ANCOVA) to determine the influence of the 3 parameters under study on the fatigue behavior ($\alpha$=.05).

**Results.** Linear regression models showed a statistical difference ($P<.001$) when the implant body diameter was increased by 1 mm; an average 3.5-fold increase in fatigue life was observed. Increasing the implant abutment connection diameter by 1.4 mm also showed a significant difference ($P<.001$), leading to 7-fold longer fatigue life on average. No significant statistical evidence was found to demonstrate a difference in fatigue life between internal and external implant-abutment connection types.

**Conclusions.** Increasing the implant platform and body diameter significantly improved ($P<.001$) the fatigue life of the prosthetic screw, whereas external and internal connections provided similar results. In addition, experimental results proved the accuracy of the fatigue life prediction methodology. (J Prosthet Dent 2022;127:477.e1-e9)

Implant-supported restorations typically consist of a dental implant, an abutment, and a prosthetic screw. The screw is the critical component of most dental restorations (except for narrow implants) with regard to mechanical failure,[1,2] especially with butt-joint implants, because it is responsible for providing structural integrity to the whole assembly by means of the preload force generated by the applied tightening torque.[3-5] A screw malfunction may lead to an inadequate assembly joint, with large abutment-implant microgaps,[6,7] self-loosening,[5,8-12] and, occasionally, the loss of the whole restoration due to fatigue failure.[3,8,13-20] Thus, knowledge about how geometric parameters of dental implants affect the prosthetic screw is a major concern. The most studied geometrical parameters have been implant length, body diameter, platform diameter, and implant-abutment connection (IAC) type.

The implant body diameter has been reported to be one of the most important factors in the biomechanical behavior of dental restorations, with a wider implant

Supported by the Basque Government [grant number IT947-16]. Supported by the University of the Basque Country through the project LTC AENIGME [COLAB19/04].
[a]PhD student, Department of Mechanical Engineering, University of the Basque Country, Bilbao, Spain; and Engineer, R&D Department, Biotechnology Institute I Mas D S.L., Miñano, Spain.
[b]Associate Professor, Department of Mechanical Engineering, University of the Basque Country, Bilbao, Spain.
[c]Lecturer and Researcher, Department of Mechanical Engineering, University of the Basque Country, Bilbao, Spain.
[d]Professor, Institut de Mécanique et Ingénierie de Bordeaux - I2M, École Nationale Supérieure d'Arts et Métiers ParisTech à Bordeaux-Talence, Talence Cedex, France.

> **Clinical Implications**
>
> For butt-joint implant-supported restorations where the prosthetic screw is the critical component, a larger implant platform and body diameter should significantly improve the fatigue life. The implant-abutment connection type, internal or external, should not significantly affect fatigue behavior.

being beneficial[21] because it increases the contact surface with the surrounding bone, thus improving stress distribution[22-25] and providing enhanced initial stability.[26,27] In addition, with increased implant diameter, stresses in the implant are reduced, especially around the implant neck,[28] improving the static and fatigue response of dental restorations.[29,30] However, these in vitro studies focused on implant failure rather than the behavior of the prosthetic screw. The authors are unaware of studies on the effect of implant body diameter on the mechanical behavior of the prosthetic screw.

Implant length has been a controversial topic, with some authors reporting lower success rates for short and extrashort implants,[31-35] whereas others have reported higher survival rates.[36-39] These discrepancies may be explained by the fact that short and extrashort implants are mainly used in clinical situations with reduced alveolar bone height, where the experience and skills of the clinician are critical. Nevertheless, the influence of implant length has been reported to be much lower than that of other parameters such as implant body diameter.[28,40] Regarding implant-bone interface, where an excessive strain may lead to bone loss, stress is mainly distributed along the first 6 threads of the implant,[41] the peak stress being at bone crest level.[42-44] Consequently, unnecessarily increasing the length of the implant may produce limited improvements, even though a longer implant may improve primary stability in situations where cancellous bone is predominant.[43,45]

The diameter of the implant-abutment platform is also a key geometrical parameter both clinically and mechanically. From a clinical point of view, the reduction of the contact diameter of the IAC is widely used in the platform switching concept,[46] where an abutment narrower than the implant is used. Platform switching can lead to reduced peri-implant bone loss.[47] Nevertheless, mechanical behavior is negatively affected by the reduction of the contact diameter of the IAC because the platform plays a primary role in joint strength, joint stability, and rotational and locational stability.[48] As reported by Minatel et al,[47] a reduction of the IAC diameter (by using the platform switching concept) can result in higher stresses in the retaining screw. In addition, Nicolas-Silvente et al[49] performed experimental fatigue tests where the retaining screws were the failing components and concluded that a higher fatigue limit was obtained with wider platforms, even though this conclusion might be limited by the fact that the specimens tested had different connection types.

External and internal butt-joint connections have been compared.[6,50-54] From a clinical point of view, internal butt-joint connections improve sealing against microbial ingress[6] and esthetics and provide more platform switching options.[50] From a mechanical point of view, IAC type may determine not only the maximum load of the restoration but also its failure mode.[51] Thus, finite element analysis (FEA) and experimental studies have determined that internal butt-joint connections have better fatigue performance than external connections.[52-54] However, these studies focused on implant failure rather than prosthetic screw failure.

Nevertheless, focusing on the mechanical behavior of the restoration, most of these studies set aside the influence of these parameters on the mechanical behavior of the prosthetic screw, which is often the critical component of the restoration when quasistatic overload or fatigue failure occurs. Moreover, the importance of the screw must not be underestimated because it may work as a mechanical fuse[51] as it is an easily replaceable component whose eventual failure secures the implant and the surrounding structure from bending overload.

The fatigue behavior of the prosthetic screw has been studied in 4 different butt-joint implant-supported restoration designs. The research hypothesis was that differences in fatigue life would be found when the implant body and platform diameters were increased and when IAC was shifted between internal and external. Implant length was not included in this study on the assumption that its effect would be negligible in comparison with the parameters studied. In addition, the authors had previously developed a theoretical fatigue life prediction methodology for prosthetic screws.[19] Therefore, the experimental test results were compared with the theoretical ones to determine the accuracy of the methodology for the wide range of designs under study.

## MATERIAL AND METHODS

Four different directly attached implant-supported restorations (BTI Biotechnology Institute) were tested (Table 1). As the study focused on prosthetic screw failures, dental restorations with narrow implants were not considered because the implant would be expected to be the failing component. All implants and abutments were made of commercially pure-grade 4 titanium (Ti CP4), and all the prosthetic screws were made of Ti6Al4V extralow interstitials (ELI) (Ti Gr 5) with the chemical composition provided in Table 2. Comparing couples of restorations as illustrated in Figure 1, the effect of implant

Table 1. Components and main parameters of restorations under study

| Restoration | IN-I4.5-P4.1 | IN-I5.5-P4.1 | IN-I5.5-P5.5 | EX-I4.5-P4.1 |
|---|---|---|---|---|
| Implant | IIPSCA4513 | IIPSCA5513 | IIPACA5513 | IRPS4513 |
| Abutment | INPPTU44 | INPPTU44 | INPPTA54 | PPTU44 |
| Screw | INTTUH | INTTUH | INTTUH | TTUH |
| Body Ø (mm) | 4.5 | 5.5 | 5.5 | 4.5 |
| Platform Ø (mm) | 4.1 | 4.1 | 5.5 | 4.1 |
| IAC | Internal | Internal | Internal | External |
| Screw Metric | M1.8 | M1.8 | M1.8 | M2 |
| Tightening torque (Ncm) | 35 | 35 | 35 | 35 |

Table 2. Chemical composition of materials used in implant and prosthetic screw components

| Ti 6Al 4V ELI (Ti GR5) | | Ti CP4 | |
|---|---|---|---|
| Composition | Wt. % | Composition | Wt. % |
| Al | 5.5-6.5 | N (max) | 0.05 |
| V | 3.5-4.5 | C (max) | 0.08 |
| Fe (max) | 0.25 | Fe (max) | 0.5 |
| O (max) | 0.13 | O (max) | 0.4 |
| C (max) | 0.08 | H (max) | 0.0125 |
| N (max) | 0.05 | - | - |
| H (max) | 0.012 | - | - |

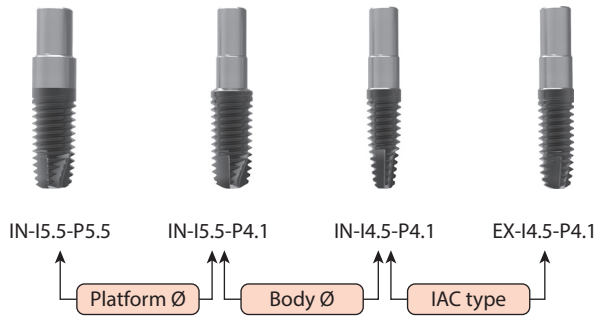

Figure 1. Dental restorations and comparisons. IAC, implant-abutment connection.

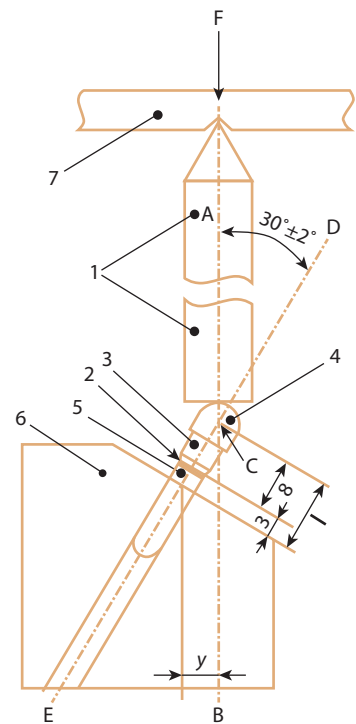

Figure 2. Schematic of test design according to ISO 14801 standard.[55] 1, Loading device. 2, Nominal bone level. 3, Implant abutment. 4, Hemispherical loading member. 5, Implant body. 6, Specimen holder. 7, Force Application. ISO, International Organization for Standardization.

body diameter, implant platform diameter, and IAC type were analyzed.

Fatigue tests were carried out in an axial fatigue testing machine (E 3000 Electropuls; Instron) with a ±5-kN load range load cell (DYNACELL 2527-153; Instron). Tests were performed according to the International Organization for Standardization (ISO) 14801 Standard with a loading ratio of 0.1.[55] Thus, each implant was placed in a specimen holder (inclined 30 degrees) extending 3 mm. The load was applied at 8 mm from the implant-abutment platform. The load varied from maximum to 10% of maximum with a frequency of 15 Hz. The load was transmitted from the actuation of the test bench to the dental restoration by means of a hemispherical device (Fig. 2). In total, 48 fatigue tests were performed for the 4 dental restorations seen in Figure 1. For each dental restoration, 3 or 4 specimens were tested at each of the 3 to 5 load levels fulfilling ISO 14801 sample size requirements[55] and at similar load ranges, covering a wide life range. Some test results for IN-I4.5-P4.1 have been published previously,[19] and additional tests (of the same manufacturing batch) were performed in the present study. All the specimens were preloaded to 35 Ncm, as recommended by the manufacturer. Fatigue test results were plotted in an F-N diagram, which relates the applied force F with the number of cycles to fatigue failure N, as indicated in ISO 14801.[55] Regression models, confidence bands for the linear regression, and prediction intervals for 95% confidence were then calculated according to the American Society of Testing and Materials (ASTM) E-739 Standard.[56]

ANCOVA calculated in a spreadsheet (Excel; Microsoft Corp) was used to statistically compare all the restorations in pairs to isolate 1 variable in each comparison (Fig. 1). In brief, the first null hypothesis assumed the same slope for both linear regression models and, if accepted, the second null hypothesis assumed the same fatigue behavior (negligible statistical differences between 2 linear regressions).

Furthermore, in previous work, a theoretical fatigue life prediction methodology for prosthetic screws in dental restorations was presented.[19] Essentially, the methodology consisted of a simple FEA that simulated

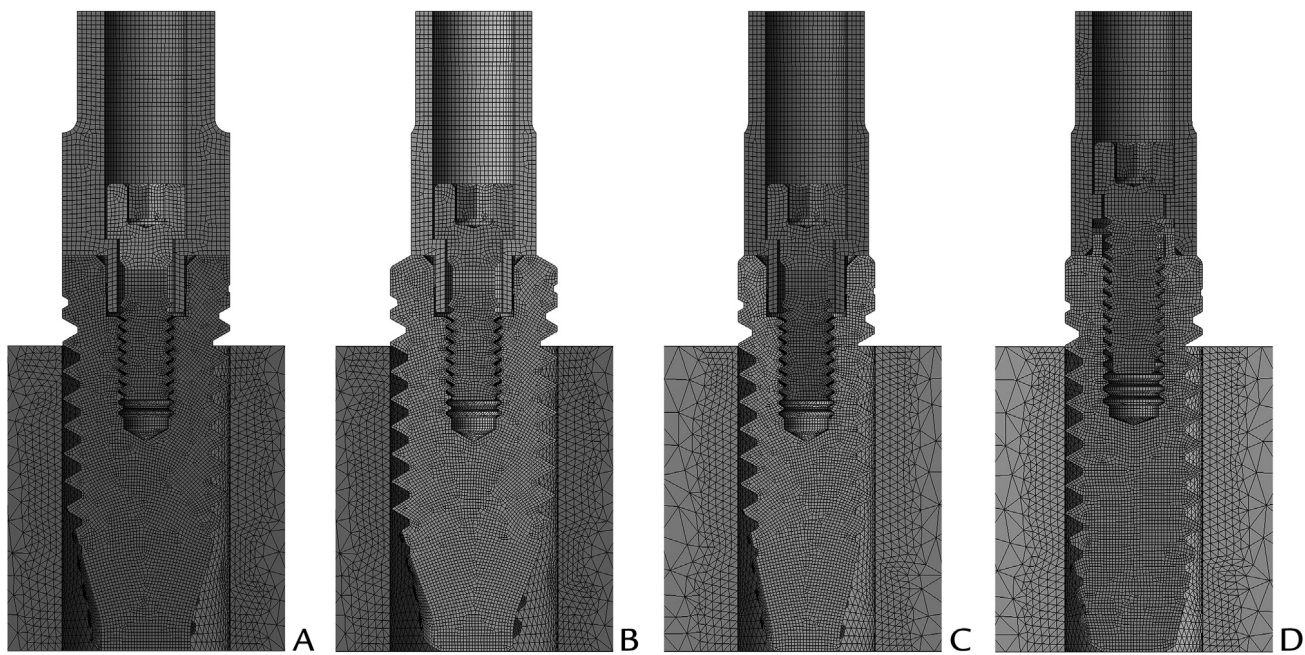

**Figure 3.** Mesh of FE models under study. A, IN-I5.5-P5.5 (2 365 587 DoF). B, IN-I5.5-P4.1 (2 202 267 DoF). C, IN-I4.5-P4.1 (1 818 912 DoF). D, EX-I4.5-P4.1 (1 671 387 DoF). DoF, degree of freedom; FE, finite element.

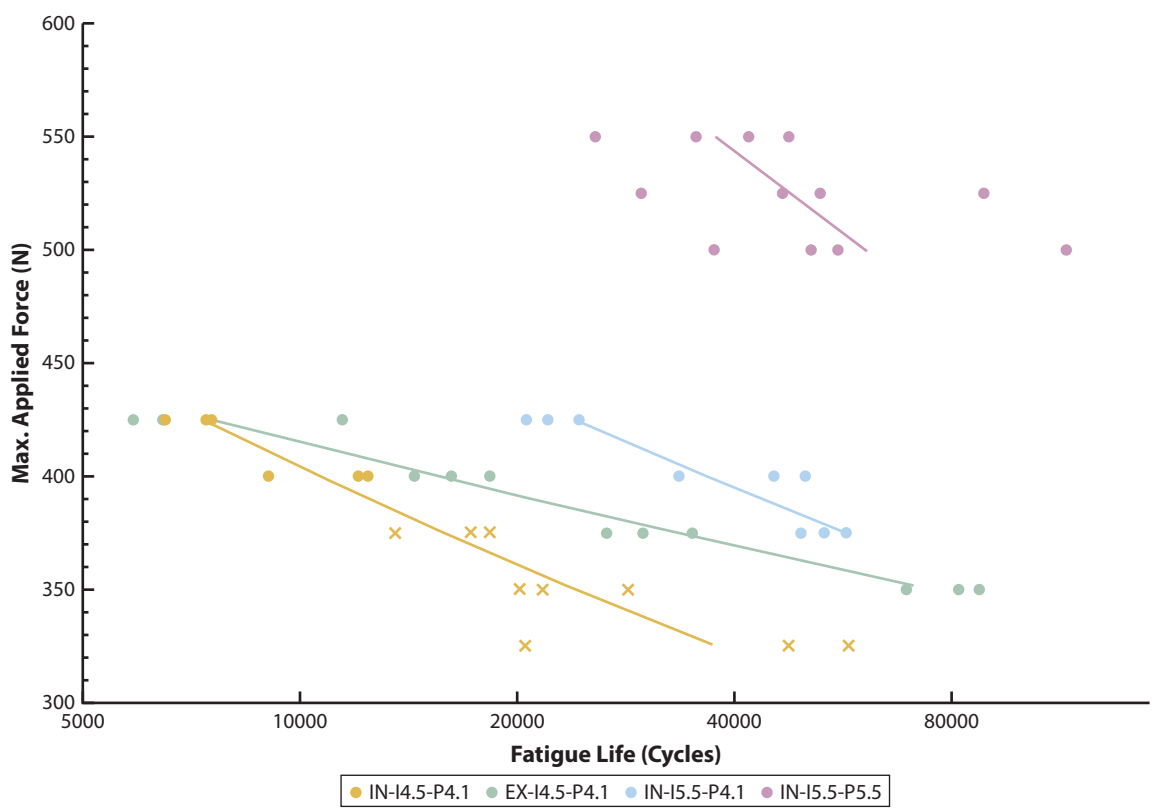

**Figure 4.** F-N curves of dental restorations under study: experimental tests (*marks*) and linear model (*lines*). Data points marked with X from previous study.[19]

ISO 14801[55] test conditions (as in the experimental test), combined with simple formulation. In this case, half geometry was modeled, and cylindrical threads were assumed in the screwed joint. Figure 3 shows the mesh of the FE models in Figure 1. From the FEA, force reactions (axial force and bending moment) in the screw head contact were considered instead of the resulting stress status, which was influenced by the mesh size and the notch effect. Thus, nominal stresses were calculated in the failure section of the screw by using well-known classical formulation from the theory of elasticity. Finally, these nominal stresses were associated with a given fatigue life by using the classical Walker fatigue equation.[57,58] Additional detailed explanation can be found in previous work.[19]

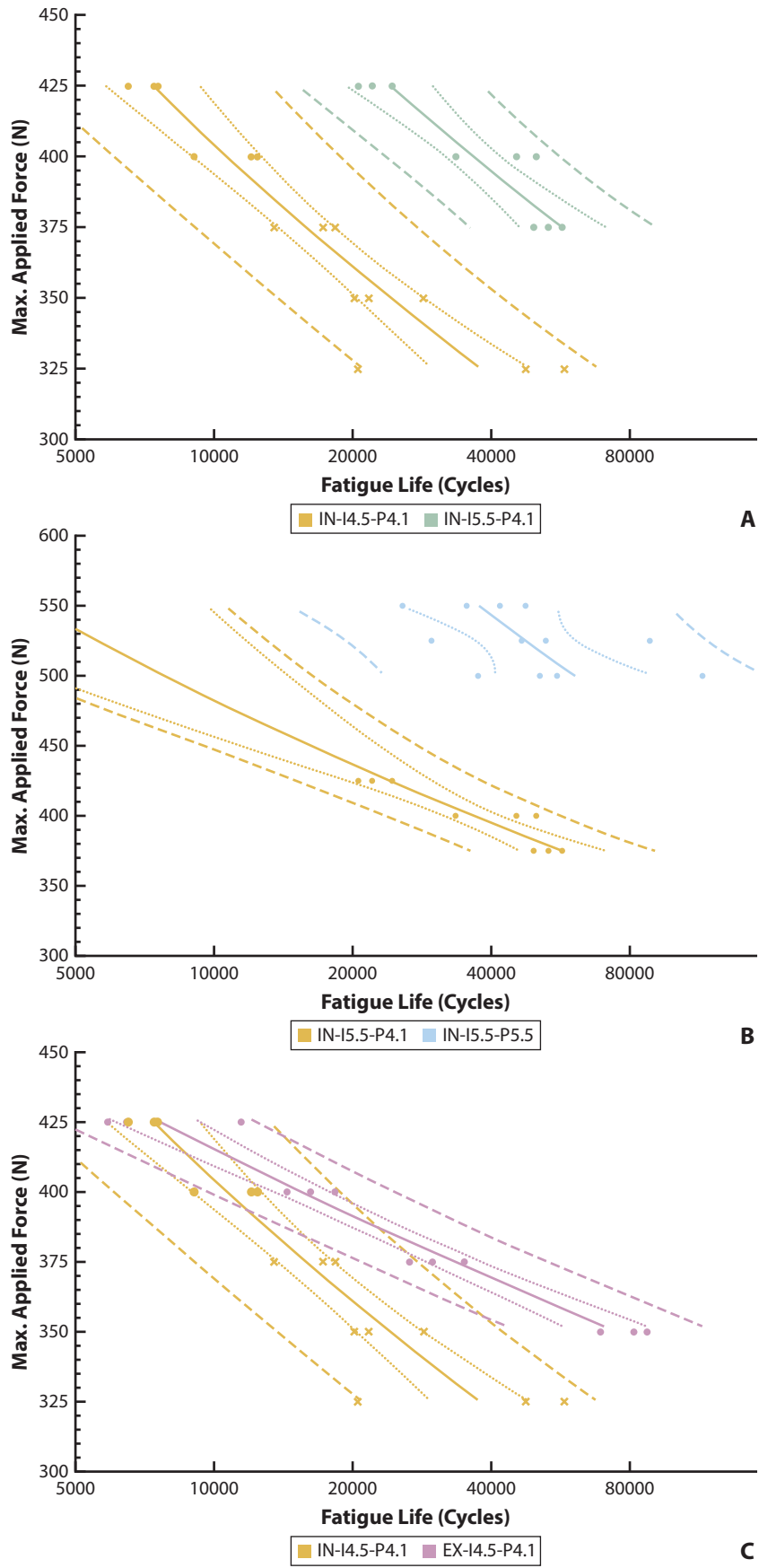

**Figure 5.** Experimental fatigue data points (*marks*), F-N curves (*lines*), linear regression confidence bands (*dotted lines*), and prediction intervals (*dashed lines*) for comparison purposes (see Fig. 1). A, Body diameter. B, Platform diameter. C, IAC type. Data points marked with X from previous study.[19] IAC, implant-abutment connection.

## RESULTS

Figure 4 shows the experimental results F-log(cycles) for each restoration, with the number of cycles to fatigue failure N for each fatigue load level F tested, and their linear models according to ISO 14801.[55] Aiming to test similar load ranges, IN-I4.5-P4.1, IN-I5.5-P4.1, and EX-I4.5-P4.1 were tested by decreasing the load by 25 N starting from 425 N; the lower limits were set in 325 N, 375 N, and 350 N, respectively, below which run-outs took place (set to 5 million cycles). IN-I5.5-P5.5 was tested in a higher load range because run-outs occurred below 500 N.

The restorations were compared in pairs (Fig. 5) to perform the comparisons presented in Figure 1. For such purpose, 95% linear regression confidence bands and 95% prediction intervals were added to the experimental points and linear models.[56] Thus, Figure 5A compares IN-I4.5-P4.1 and IN-I5.5-P4.1 to study the effect of implant body diameter; Figure 5B compares IN-I5.5-P4.1 and IN-I5.5-P5.5 for the influence of platform diameter; and, in Figure 5C, IN-I4.5-P4.1 and EX-I4.5-P4.1 were plotted to evaluate the effect of the IAC type. In all the cases, fatigue failure occurred in the first engaged thread of the prosthetic screw, as expected (Fig. 6).[59-61]

Regarding Figure 5A, ANCOVA was used to compare both linear models, accepting the first null hypothesis that the slopes were equal (P=.615) and rejecting the second null hypothesis (P<.001), that is, the mean fatigue life was statistically different. Furthermore, once the slopes of both models were determined to be equal, the fatigue life was calculated to be enhanced 3.5-fold when the implant diameter was increased from 4.5 to 5.5 mm.

IN-I5.5-P4.1 and IN-I5.5-P5.5 could not be tested in the same load range (Fig. 5B). Thus, the linear model of IN-I5.5-P4.1 was extrapolated to allow a comparison of both models. ANCOVA was used to compare the linear models, accepting the first null hypothesis that the slopes were equal (P=.541) and rejecting the second null hypothesis (P<.001); moreover, the fatigue life was calculated to be 7 times larger when the platform diameter was increased from 4.1 to 5.5 mm. This value resulted from a linear extrapolation, and this factor may be even larger because of the possible nonlinear behavior of the material when approaching the low-cycle-fatigue domain.

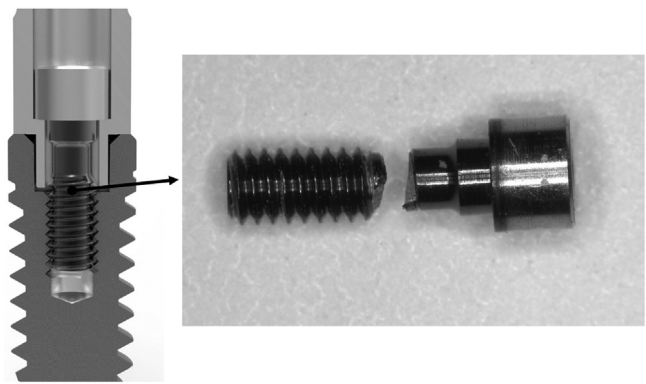

Figure 6. Fatigue failure of prosthetic screw in first engaged thread.

Finally, ANCOVA rejected the hypothesis that the slopes were equal (P<.001) for the models compared in Figure 5C, and therefore, a difference in fatigue life was not demonstrated. The linear regression confidence bands and prediction intervals for 3 of the 4 load levels overlapped, indicating that significant differences were not demonstrated in terms of fatigue life.

Figure 7 illustrates all the experimental test results (vertical axis) versus the fatigue life estimated by the theoretical prediction methodology (horizontal axis). The 45-degree black line represents a perfect tool-experimental match (experimental life equal to theoretically predicted life). However, a perfect correlation is virtually impossible because of the inherent dispersion of the fatigue phenomenon,[62] as well as the scatter of the torque-to-preload ratio in screwed joints.[12]

## DISCUSSION

The experimental fatigue test results determined that fatigue response of the prosthetic screw significantly improved when the implant body diameter was increased from 4.5 to 5.5 mm. Linear regression models showed a fatigue life 3.5 times higher for the highest load range under study, accepting the research hypothesis concerning implant body diameter. This beneficial effect was consistent with the findings of Shemtov-Yona et al[29] and Fan et al,[30] who also reported improved fatigue response with larger implant diameters. This, together with the reduction of the surrounding bone stresses[22-25] and the enhancement of the initial stability,[26,27] determined that increasing the body diameter will improve the overall behavior of the dental restoration. However, even though a large implant body diameter is recommended from a mechanical point of view, horizontal crestal bone atrophy and restricted edentulous area often limit the use of large diameters.[63] Furthermore, narrower implants decrease

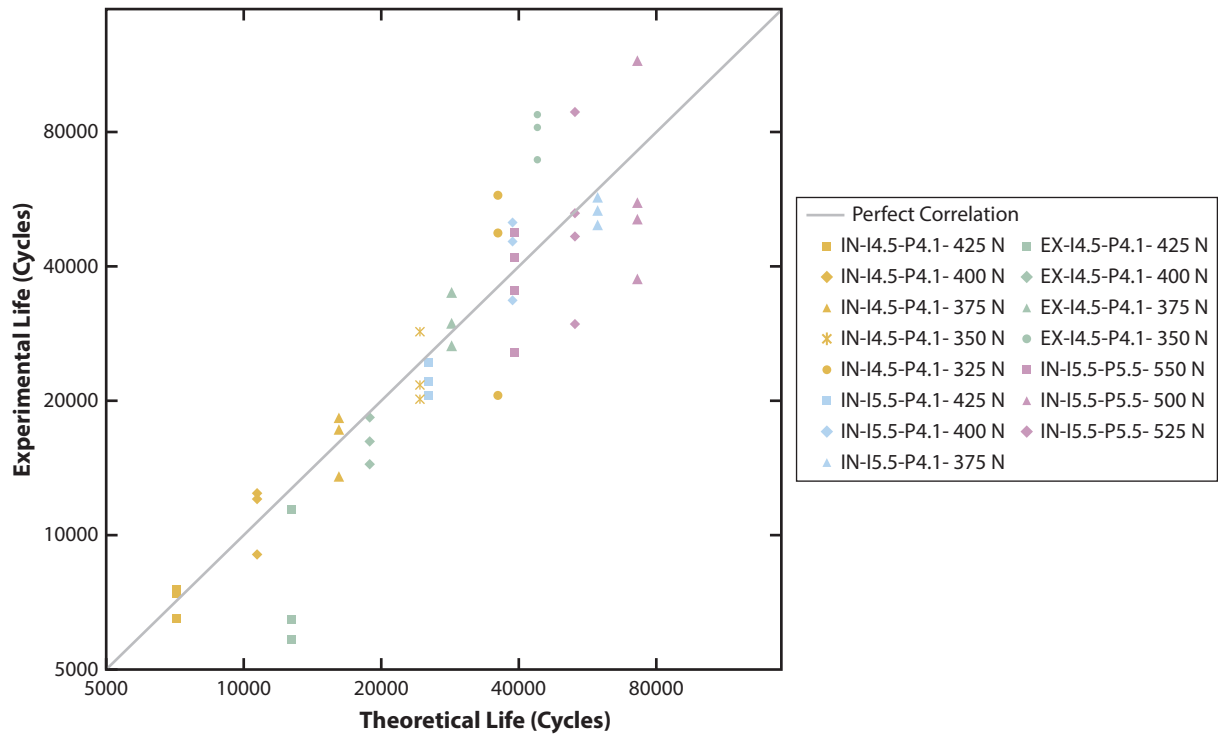

**Figure 7.** Experimental results versus theoretical life prediction.[19]

the need for bone augmentations, reducing surgical invasiveness.[64]

Regarding implant platform diameter, a major improvement in fatigue life was achieved when the platform diameter was increased from 4.1 to 5.5 mm. The number of cycles was increased by a factor of approximately 7 according to the linear models, accepting the research hypothesis concerning implant platform diameter. These results were consistent with those of Nicolas-Silvente et al,[49] who reported lower stresses on the prosthetic screw with larger platform diameters. Nevertheless, the platform switching effect on peri-implant bone should be taken into account to avoid peri-implant bone loss.[46,47]

Concerning IAC type, the results did not show a significant difference between internal and external connections, rejecting the research hypothesis of this study concerning IAC type. Nevertheless, the internal connection tested had a narrower screwed joint than the external one (M1.8 versus M2, as indicated in Table 1). Thus, a slightly improved fatigue response of the internal connections may be expected over the external connection if compared under the same conditions, consistent with previous studies.[52-54] Nevertheless, the effect of the IAC type on the fatigue life of the prosthetic screw was significantly smaller than the other 2 parameters studied, and internal butt-joint connections have other advantages, including improved esthetics, sealing, and platform switching options.[6,50]

The fatigue life prediction methodology previously developed by the authors[19] provided accurate life predictions for the wide range of geometric dimensions studied. The authors believe that this methodology can be used by manufacturers to compare different implant designs, including different geometries, as in the present study, loading rates, tightening torques, and friction coefficients. Currently these comparisons are carried out by building and experimentally testing prototypes, a costly and time-consuming process. The theoretical methodology is a powerful, versatile, and cost-effective design tool. The methodology predicts the fatigue life for the dental restoration when the prosthetic screw is the critical component, so it cannot be used for very narrow dental restorations where the implant is the failing component. For other materials and coating or surface treatments, the parameters of the fatigue life prediction equation must be tuned as explained previously.[19]

## CONCLUSIONS

Based on the findings of this in vitro study, the following conclusions were drawn:

1. Increasing the implant body diameter, and especially the implant platform diameter, increased the fatigue life of the prosthetic screw.
2. No significant difference was found in fatigue life between internal and external IAC types.

3. The theoretical fatigue life predictions were consistent with the experimental results.

## REFERENCES

1. Dhima M, Paulusova V, Lohse C, Salinas TJ, Carr AB. Practice-based evidence from 29-year outcome analysis of management of the edentulous jaw using osseointegrated dental implants. J Prosthodont 2014;23:173-81.
2. Pjetursson B, Asgeirsson A, Zwahlen M, Sailer I. Improvements in implant dentistry over the last decade: comparison of survival and complication rates in older and newer publications. Int J Oral Maxillofac Implants 2014;29: 308-24.
3. Stüker RA, Teixeira ER, Beck JCP, Da Costa NP. Preload and torque removal evaluation of three different abutment screws for single standing implant restorations. J Appl Oral Sci 2008;16:55-8.
4. Barbosa GS, da Silva-Neto JP, Simamoto-Júnior PC, das Neves FD, da Gloria Chiarello de Mattos M, Ribeiro RF. Evaluation of screw loosening on new abutment screws and after successive tightening. Braz Dent J 2011;22:51-5.
5. Lang LA, Kang B, Wang RF, Lang BR. Finite element analysis to determine implant preload. J Prosthet Dent 2003;90:539-46.
6. Mishra SK, Chowdhary R, Kumari S. Microleakage at the different implant abutment interface: a systematic review. J Clin Diagn Res 2017;11:ZE10-5.
7. Aguirrebeitia J, Abasolo M, Vallejo J, Ansola R. Dental implants with conical implant-abutment interface: influence of the conical angle difference on the mechanical behavior of the implant. Int J Oral Maxillofac Implants 2013;28: e72-82.
8. Schwarz MS. Mechanical complications of dental implants. Clin Oral Implants Res 2000;11:156-8.
9. Wu T, Fan H, Ma R, Chen H, Li Z, Yu H. Effect of lubricant on the reliability of dental implant abutment screw joint: an in vitro laboratory and three-dimension finite element analysis. Mater Sci Eng C 2017;75:297-304.
10. Elias CN, Figueira DC, Rios PR. Influence of the coating material on the loosing of dental implant abutment screw joints. Mater Sci Eng C 2006;26: 1361-6.
11. Siamos G, Winkler S, Boberick KG. The relationship between implant preload and screw loosening on implant-supported prostheses. J Oral Implantol 2002;28:67-73.
12. Armentia M, Abasolo M, Coria I, Bouzid A-H. On the use of a simplified slip limit equation to predict screw self-loosening of dental implants subjected to external cycling loading. Appl Sci 2020;10:6748.
13. Bickford JH. Introduction to the design and behavior of bolted joints. 4th ed. New York: CRC Press; 2008. p. 1-14.
14. Zeno HA, Buitrago RL, Sternberger SS, Patt ME, Tovar N, Coelho P, et al. The effect of tissue entrapment on screw loosening at the implant/abutment interface of external- and internal-connection implants: an in vitro study. J Prosthodont 2016;25:216-23.
15. Jeng M-D, Lin Y-S, Lin C-L. Biomechanical evaluation of the effects of implant neck wall thickness and abutment screw size: a 3D nonlinear finite element analysis. Appl Sci 2020;10:3471.
16. Jeng MD, Liu PY, Kuo JH, Lin CL. Load fatigue performance evaluation on two internal tapered abutment-implant connection implants under different screw tightening torques. J Oral Implantol 2017;43:107-13.
17. Quek H, Tan K, Nicholls J. Load fatigue performance of four implant-abutment interface designs: effect of torque level and implant system. J Prosthet Dent 2008;23:253-62.
18. Xia D, Lin H, Yuan S, Bai W, Zheng G. Dynamic fatigue performance of implant-abutment assemblies with different tightening torque values. Bio-med Mater Eng 2014;24:2143-9.
19. Armentia M, Abasolo M, Coria I, Albizuri J. Fatigue design of dental implant assemblies: a nominal stress approach. Metals (Basel) 2020;10:744.
20. Abasolo M, Aguirrebeitia J, Vallejo J, Albizuri J, Coria I. Influence of vertical misfit in screw fatigue behavior in dental implants: a three-dimensional finite element approach. Proc Inst Mech Eng Part H J Eng Med 2018;232:1117-28.
21. Kang N, Wu YY, Gong P, Yue L, Ou GM. A study of force distribution of loading stresses on implant-bone interface on short implant length using 3-dimensional finite element analysis. Oral Surg Oral Med Oral Pathol Oral Radiol 2014;118:519-23.
22. Santiago Junior JF, Pellizzer EP, Verri FR, De Carvalho PSP. Stress analysis in bone tissue around single implants with different diameters and veneering materials: a 3-D finite element study. Mater Sci Eng C 2013;3: 4700-14.
23. Chang SH, Lin CL, Hsue SS, Lin YS, Huang SR. Biomechanical analysis of the effects of implant diameter and bone quality in short implants placed in the atrophic posterior maxilla. Med Eng Phys 2012;34:153-60.
24. Raaj G, Manimaran P, Kumar C, Sadan D, Abirami M. Comparative evaluation of implant designs: influence of diameter, length, and taper on stress and strain in the mandibular segment-a three-dimensional finite element analysis. J Pharm Bioallied Sci 2019;16:486-94.
25. Langer B, Langer L, Herrmann I, Jorneus L. The wide fixture: a solution for special bone situations and a rescue for the compromised implant. Part 1. Int J Oral Maxillofac Implants 1993;8:400-8.
26. Lee JH, Frias V, Lee KW, Wright RF. Effect of implant size and shape on implant success rates: a literature review. J Prosthet Dent 2005;94:377-81.
27. Misch CE. Implant design considerations for the posterior regions of the mouth. Implant Dent 1999;8:376-86.
28. Himmlová L, Dostálová T, Kácovský A, Konvičková S. Influence of implant length and diameter on stress distribution: a finite element analysis. J Prosthet Dent 2004;91:20-5.
29. Shemtov-Yona K, Rittel D, Levin L, Machtei EE. Effect of dental implant diameter on fatigue performance. Part I: mechanical behavior. Clin Implant Dent Relat Res 2014;16:172-7.
30. Fan H, Gan X, Zhu Z. Evaluation of dental implant fatigue performance under loading conditions in two kinds of physiological environment. Int J Clin Exp Med 2017;10:6369-77.
31. Horiuchi K, Uchida H, Yamamoto K, Sugimura M. Immediate loading of Branemark system implants following placement in edentulous patients: a clinical report. Int J Oral Maxillofac Implants 2000;15:824-30.
32. Chiapasco M, Abati S, Romeo E, Vogel G. Implant-retained mandibular overdentures with Branemark system MKII implants: a prospective comparative study between delayed and immediate loading. Int J Oral Maxillofac Implants 2001;16:537-46.
33. Bakaeen LG, Winkler S, Neff PA. The effect of implant diameter, restoration design, and occlusal table variations on screw loosening of posterior single-tooth implant restorations. J Oral Implantol 2001;17:63-72.
34. Winkler S, Morris HF, Ochi S. Implant survival to 36 months as related to length and diameter. Ann Periodontol 2000;5:22-31.
35. Wyatt C, Zarb G. Treatment outcomes of patients with implant-supported fixed partial prostheses. Int J Oral Maxillofac Implants 1998;13:204-11.
36. Anitua E, Alkhraisat MH. 15-year follow-up of short dental implants placed in the partially edentulous patient: mandible vs maxilla. Ann Anat 2019;222: 88-93.
37. Anitua E, Piñas L, Orive G. Retrospective study of short and extra-short implants placed in posterior regions: influence of crown-to-implant ratio on marginal bone loss. Clin Implant Dent Relat Res 2015;17:102-10.
38. Anitua E, Piñas L, Begoña L, Orive G. Long-term retrospective evaluation of short implants in the posterior areas: clinical results after 10-12 years. J Clin Periodontol 2014;41:404-11.
39. Anitua E, Orive G. Short implants in maxillae and mandibles: a retrospective study with 1 to 8 years of follow-up. J Periodontol 2010;81:819-26.
40. Topkaya H, Kaman MO. Effect of dental implant dimensions on fatigue behaviour: a numerical approach. Uludag Univ J Fac Eng 2018;23:249-60.
41. Anitua E, Tapia R, Luzuriaga F, Orive G. Influence of implant length, diameter, and geometry on stress distribution: a finite element analysis. Int J Periodontics Restorative Dent 2010;30:89-95.
42. Georgiopoulos B, Kalioras K, Provatidis C, Manda M, Koidis P. The effects of implant length and diameter prior to and after osseointegration: a 2-D finite element analysis. J Oral Implantol 2007;33:243-56.
43. Baggi L, Cappelloni I, Di Girolamo M, Maceri F, Vairo G. The influence of implant diameter and length on stress distribution of osseointegrated implants related to crestal bone geometry: a three-dimensional finite element analysis. J Prosthet Dent 2008;100:422-31.
44. Lum LB. A biomechanical rationale for the use of short implants. J Oral Implantol 1991;17:126-31.
45. Hingsammer L, Pommer B, Hunger S, Stehrer R, Watzek G, Insua A. Influence of implant length and associated parameters upon biomechanical forces in finite element analyses: a systematic review. Implant Dent 2019;28: 296-305.
46. Atieh MA, Ibrahim HM, Atieh AH. Platform switching for marginal bone preservation around dental implants: a systematic review and meta-analysis. J Periodontol 2010;81:1350-66.
47. Minatel L, Verri FR, Kudo GAH, de Faria Almeida DA, de Souza Batista VE, Lemos CAA, et al. Effect of different types of prosthetic platforms on stress-distribution in dental implant-supported prostheses. Mater Sci Eng C 2017;71:35-42.
48. Shadid R, Sadaqah N, Al-Omari W, Abu-Naba'a L. Comparison between the butt-joint and morse taper implant-abutment connection: a literature review. J Implant Adv Clin Dent 2013;5:33-40.
49. Nicolas-Silvente AI, Velasco-Ortega E, Ortiz-Garcia I, Jimenez-Guerra A, Monsalve-Guil L, Ayuso-Montero R, et al. Influence of connection type and platform diameter on titanium dental implants fatigue: non-axial loading cyclic test analysis. Int J Environ Res Public Health 2020;17: 8988.
50. Shafie HR. Clinical and laboratory manual of dental implant abutments. New Jersey: John Wiley & Sons, Inc; 2014. p. 33-46.
51. Dittmar S, Dittmer MP, Kohorst P, Jendras M, Borchers L, Stiesch M. Effect of implant-abutment connection design on load bearing capacity and failure mode of implants. J Prosthodont 2011;20:510-6.
52. Yamaguchi S, Yamanishi Y, Machado LS, Matsumoto S, Tovar N, Coelho PG, et al. In vitro fatigue tests and in silico finite element analysis of dental implants with different fixture/abutment joint types using computer-aided design models. J Prosthodont Res 2018;62:24-30.
53. Khraisat A, Stegaroiu R, Nomura S, Miyakawa O. Fatigue resistance of two implant/abutment joint designs. J Prosthet Dent 2002;88:604-10.
54. Fernández-Asián, Martínez-González, Torres-Lagares, Serrera-Figallo, Gutiérrez-Pérez. External connection versus internal connection in dental implantology. A mechanical in vitro study. Metals (Basel) 2019;9:1106.
55. International Organization for Standardization. ISO 14801. Dentistry. Implants. Dynamic fatigue test for endosseous dental implants. 3rd. Geneva: International Organization for Standardization; 2016. p. 1-10.
56. ASTM International. ASTM E739-91. Standard practice for statistical analysis of linear or linearized stress-life (S–N) and strain-life (e–N) fatigue data. West Conshohocken: ASTM International; 2004. p. 1-6.
57. Walker K. The effect of stress ratio during crack propagation and fatigue for 2024-T3 and 7075-T6 aluminum. In: Effects of environment and complex load history on fatigue life. West Conshohocken: ASTM International; 1970. p. 1-14.
58. Dowling NE, Calhoun CA, Arcari A. Mean stress effects in stress-life fatigue and the Walker equation. Fatigue Fract Eng Mater Struct 2009;32:163-79.
59. Coria I, Abasolo M, Gutiérrez A, Aguirrebeitia J. Achieving uniform thread load distribution in bolted joints using different pitch values. Mech Ind 2020;21:616.
60. Pérez MA. Life prediction of different commercial dental implants as influence by uncertainties in their fatigue material properties and loading conditions. Comput Methods Programs Biomed 2012;108:1277-86.
61. Yamanaka S, Amiya K, Saotome Y. Effects of residual stress on elastic plastic behavior of metallic glass bolts formed by cold thread rolling. J Mater Process Technol 2014;214:2593-9.
62. Schijve J. Fatigue and scatter. In: Schijve J, editor. Fatigue of structures and materials. Dordrescht: Springer Netherlands; 2009. p. 373-94.
63. Chiapasco M, Casentini P, Zaniboni M, Corsi E, Anello T. Titanium-zirconium alloy narrow-diameter implants (Straumann Roxolid®) for the rehabilitation of horizontally deficient edentulous ridges: prospective study on 18 consecutive patients. Clin Oral Implants Res 2012;23:1136-41.
64. Klein M, Schiegnitz E, Al-Nawas B. Systematic review on success of narrow-diameter dental implants. Int J Oral Maxillofac Implants 2014;29:43-54.

**Corresponding author:**
Dr Mikel Armentia
Department of Mechanical Engineering
University of the Basque Country (UPV/EHU)
Plaza Ingeniero Torres Quevedo, 1
Bilbao 48013
SPAIN
Email: marmentia002@ikasle.ehu.eus